# Development of theoretical descriptors for cytotoxicity evaluation of metallic nanoparticles

D. W. Boukhvalov* and T.H. Yoon

*Department of Chemistry, Hanyang University, 17 Haengdang-dong, Seongdong-gu, Seoul 04763 Korea*

*Motivated by the recent development of quantitative structure-activity relationship (QSAR) methods in the area of nanotoxicology, we proposed an approach to develop additional descriptors based on results of first principles calculations. For evaluation of the biochemical activity of metallic nanoparticles, we consider two processes: ion extraction from the surface of a specimen to aqueous media and water dissociation on the surface. We performed calculations for a set of metals (Al, Fe, Cu, Ag, Au, Pt). Taking into account the diversity of atomic structures of real metallic nanoparticles, we performed calculations for different models such as (001) and (111) surfaces, nanorods, and two different cubic nanoparticles of 0.6 and 0.3 nm size. Significant energy dependence of the processes from the selected model of nanoparticle suggests that for the correct description we should combine the calculations for the several representative models. In addition to the descriptors of chemical activity of the metallic nanoparticles for the two studied processes, we propose descriptors for taking into account the dependence of chemical activity from the size and shape of nanoparticles. Routes to minimization of computational costs for these calculations are also discussed.*

E-mail: danil@hanyang.ac.kr

Phone: +82-2220-0966

## 1. Introduction

A recent increase in the number of industrial products utilizing nanoparticles, and the detection of the presence of various objects of nanometer sizes in industrial and natural pollutions, require the development of evaluation methods of possible hazards of nanoparticles (meaning all types of nano-sized objects) to living organisms.[1,2] A large list of nanoparticles makes it impossible to study the cytotoxicity of each type of nanoparticle separately, and requires the employment of 'big data' methods for the prediction of their biochemical activity. Similar methods used in predictive drug design[3-10] predictions can be based on the analysis of several numerical characteristics of so-called nanoparticle predictors. Increasing of exactness of predictions requires extending the list of numerical characteristics (descriptors) of biochemical activity of nanoparticles. Some descriptors can be obtained from first principles calculations. The difference in the chemical structure of organic molecules used in drug design and nanoparticles makes it impossible for mechanical transfers from the set of descriptors used in QSAR (quantitative structure-activity relationship) methods[11] to predict nanotoxicology.

The main difference between organic molecules evaluated as potential drugs evaluated by QSAR methods and nanoparticles, is the uncertainty of exact atomic structures of nanoparticles. Real nanoparticles usually have different sizes that correspond to a different number of atoms in the particles. Nanoparticles of similar size and number of atoms can be significantly different in shape, that correspond with different volume/surface ratios, as well as various crystallographic surfaces (such as (001), (111), etc.) and varied amounts of topological defects of the surface (steps, adatoms etc.). Nanoparticles can be spherical, cubic, pyramidal, etc., or have rod-like or other shapes. Multiple studies (see for review refs. [12-14]) of metal and oxide catalysts demonstrate significant differences in the chemical activity of different surfaces, in addition to the colossal role of the imperfectness of the surface for their stability and chemical activity.

Recent studies of the toxic activity of nanoparticles usually only considered the chemical composition of the nanoparticles[15-22] without taking into account the aforementioned structural issues, or only use several nanoclusters for the model on the atomic level and those interactions with selected biomolecules.[23-29] In our work, we evaluate the role of the topological aspects (sizes and shapes) of metallic nanoparticles to the energetics of the processes related to biochemical activity, discuss the methods for the calculations of numerical descriptors of nanoparticles, and the minimal model for these calculations.

## 2. Choosing the proper model

The main question for modeling of biochemical activity of nanoparticles is the choice of a representative chemical process. In recent work,[15] data relevant to two chemical processes were discussed. The first process is redox activity and the second is the extraction of a metallic ion from the surface of the oxide. To evaluate the role of the features of surfaces of nanoparticles, we perform calculations of reaction enthalpy for the processes. The first process is water decomposition on the surface of nanoparticles with further chemisorption of hydrogen and hydroxyl groups on this surface (Fig. 1a, 1b). This process is representative because almost all biochemical processes occur in aqueous media and water decomposition is the minimal model for numerical characterization of redox activity of materials.

The second process is the extraction of a single ion from the pristine surface of a nanoparticle to aqueous media with the formation of a $M(H_2O)_2$ complex (Fig. 1c), and a similar process on the surface with a dissociated water molecule (see above) with the formation of a MOH(H$_2$O) complex (Fig. 1d). We will refer to these processes as oxidation and ion extraction. An additional advantage of this approach is it takes into account the role of media in the process of ion extraction, which could change the energetics of ion extraction, not only quantitatively, but also qualitatively. As discussed above, M(H$_2$O)$_2$ and MOH(H$_2$O) complexes are the minimal model that take into account extraction of the ion, not in a vacuum, but in water with further coordination of water molecules. This model underestimates the contribution the coordination of the extracted ion by several other water molecules (note that exact number of water molecules which coordinate with the ion of metal in biological systems is an unknown value), but creates a minimal reasonable coordination. Considering a larger number of water molecules provides additional coordination of the extracted ion without significant changes of the energetics of the extraction processes (see discussion below).

In our work, we performed modeling for several metals (M = Al, Fe, Cu, Ag, Au, Pt). The choice of compounds was determined for several reasons, such as the abundance of experimental data about cytotoxicity these metals, employment of these materials in medicine (Au, Pt), or the high probability of the presence of these metals in industrial pollution (Al, Fe, Cu). Instability in living bodies (for example, Fe and Cu nanoparticles are unstable while Au and Pt are stable) and toxic harm from the material (for example Au and Pt are safe while Ag is sometimes toxic) also guided our selection.

The building of a proper structural model of nanoparticles is the actual subject of theoretical nanoscience. The relation of the time of computation (t) and the number of atoms in the system (N) are t ~ $N^3$.[30] At the current level of hardware and software development, only calculations for systems with about 300~500 atoms can be done in realistic time. This number of atoms corresponds with nanoparticles about the size of 3 nm which is much smaller than the size of nanoparticles evaluated in toxicological experiments (above 10 nm, usually 50-1000 nm).

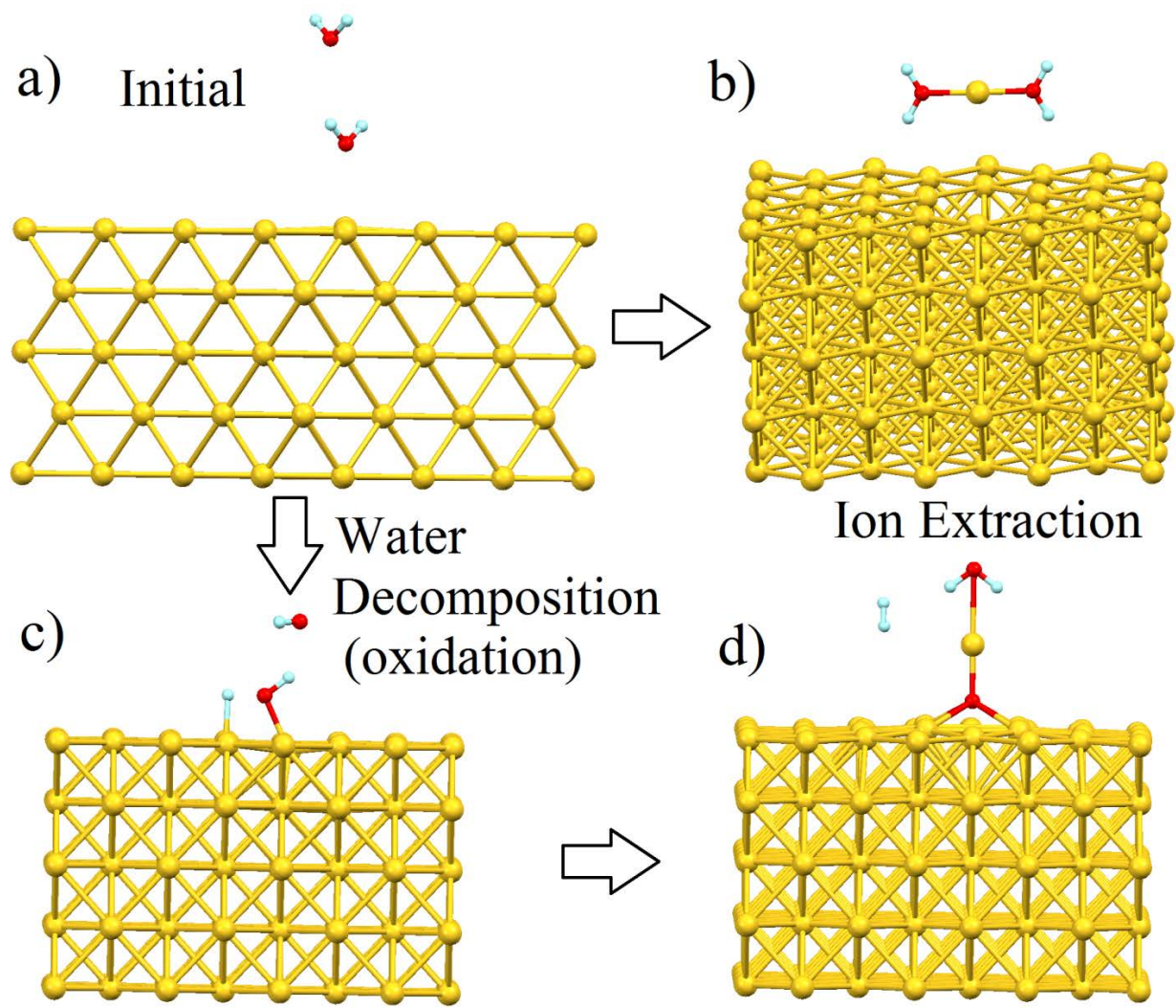

**Figure 1.** Optimized atomic structure of (001) surface of a gold slab with two physically adsorbed water molecules (a), after extraction of a gold ion with the formation of the $Au(H_2O)_2$ complex (b), after decomposition of the water molecule on the surface (c) and further extraction of the gold atom connected with the hydroxyl group and the formation of $AuOH(H_2O)$ complex.

There are two different methods to solve the problem of high computational costs. One is the use of a slab (see Fig. 1) as a feasible model for the surface of nanoparticles. This approach is reasonable because the cubic nanoparticle is 10 nm, the length of each side is about 30 lattice parameters, and the bigger part of the surface nanoparticle is more than 1 nm (about 3 lattice parameters). Because of this, atoms located far from the edges and corners have similar properties as atoms on the infinite surface. For the modeling of the surface, usually one employs a slab of at least 6 layers within periodic boundary conditions. For good separation between single processes on the surface, the size of a supercell on the xy-plane should be at least 1 nm (see Fig. 1). Because different crystallographic surfaces have different chemical activity, and all or some of them can be presented as real nanoparticles, this issue should be considered. In this work, we will compare two most propagated kinds of surfaces (111) and (001).

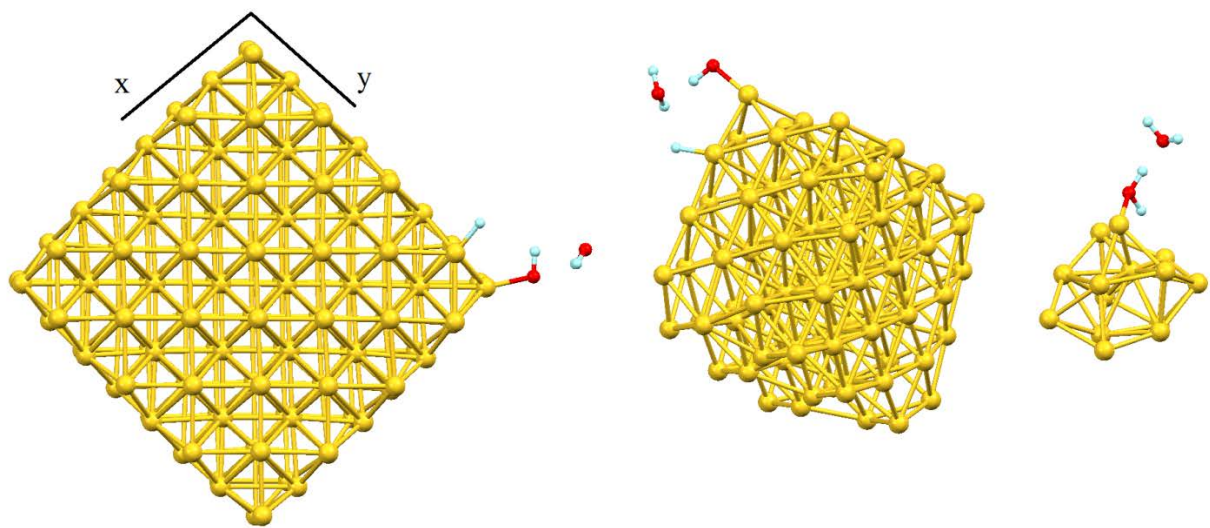

**Figure 2.** Optimized atomic structures of water decomposition over the edge of gold nanowire (infinite along the z-axis), and over the corners of gold cubic nanoparticles 0.6 and 0.3 nm size.

Usage of the slab as the model of nanoparticles excludes from consideration chemically active parts of nanoparticles such as corners, steps, and other topological features. The chemical activity of these special areas depends on not only on local topology, but also on the chemical composition of the material which requires addition calculations. For the model of these areas, we used nanowire of square-like sequence (Fig. 2a) and two cubic nanoparticles of 0.6 and 0.3 nm size (Fig. 2b, c). Because multiple experimental works[31-33] demonstrate that the crystal structure of nanowires and nanoparticles coincides with bulk materials, we also use this crystal structure as an initial set for the calculation of atomic structure of discussed nano-objects. An additional reason for performing the calculations for the smallest nanoparticle is a test of the minimal model for calculations.

## 4. Computational Method

The modeling of the adsorption of various molecules on the surface of metals is carried out by the density functional theory realized in the pseudopotential code SIESTA.[30] All calculations are done using the GGA-PBE approximation[34] while taking into account van der Waals corrections.[35] All calculations were carried out in the spin-polarized mode, for energy mesh cut off was 360 Ry and k-point mesh 3×3×1, in the Monkhorst–Pack scheme[36] for slabs 3×3×4, for nanowires, and 4×4×4 for nanoparticles. During the optimization, the electronic ground state was found self-consistently using norm-conserving pseudopotentials[37] for cores, a double-$\zeta$ plus polarization basis of localized orbitals for metals and oxygen, and a double-$\zeta$ basis for hydrogen. Optimization of bond lengths and total energies were performed with an accuracy of 0.04 eV/Å and 1 meV, respectively. The enthalpy of reaction for each process was defined as the total energy difference between the initial and final configurations.

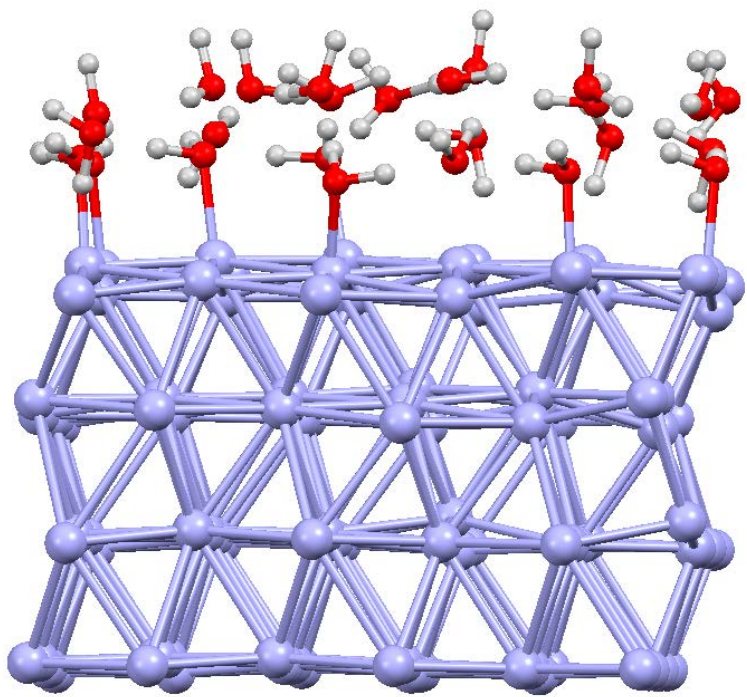

**Figure 3.** Optimized atomic structure of two layers of water over (001) surface of aluminum.

For the modeling of the activity of large nanoparticles we used slabs of 5 layers (see Fig. 1) within periodic boundary conditions. The number of layers corresponds with the minimal number of internal layers that is greater than the number of surface layers. The separation distance between the slabs is 6 lattice parameters that correspond with values in the order of 2 nm. We also take into account dipole correction.[38] For the simulation of nanowires, we use a cementite supercell (see Fig. 2) within periodical boundary conditions along the z axis and separated by 2 nm of empty space between nanowires. For the modelling of nanoparticles, we start from a bulk-like cubic configuration within an empty cubic box of 4 nm size. To check for the account solvent effect, we also perform calculations for (001) surface of aluminum covered by two layers of water molecules (Fig. 3).

**Table 1.** Calculated enthalpy of reaction ("energy cost", in eV) of extraction of ion before water dissociation (Fig. 1b), water dissociation (Fig. 1c), and extraction of the ion after water dissociation (Fig. 1d) for two types of surfaces, nanowire, and two types of nanoparticles. For the aluminum (001) surface in parenthesis, we report results for the complete coverage of the surface by water (see Fig. 3) and discussion in the text.

| Material | (001) surface | (111) surface | Nanowire d ~1.1 nm | Nanoparticle-1 d ~ 0.6 nm | Nanoparticle-2 d ~ 0.3 nm |
|---|---|---|---|---|---|
| Ion extraction (Fig. 1b) | | | | | |
| Al | 4.87 (4.23) | 0.25 | 1.54 | 0.89 | 0.45 |
| Fe | 1.99 | 2.41 | 4.49 | -1.21 | -3.12 |
| Cu | 3.04 | 6.78 | 1.92 | 3.63 | 0.35 |
| Ag | 4.22 | 1.90 | 3.77 | 1.88 | 0.90 |
| Au | 4.75 | 2.66 | 3.93 | 2.34 | 0.53 |
| Pt | 5.02 | 2.48 | 5.97 | 4.38 | 1.27 |
| Water decomposition (oxidation, Fig. 1c) | | | | | |
| Al | 0.21 (0.25) | 0.99 | -2.48 | -0.44 | -3.51 |
| Fe | -1.88 | -2.46 | -0.32 | -2.77 | -6.11 |
| Cu | 0.44 | 1.09 | -1.30 | -1.66 | -0.21 |
| Ag | 0.59 | 0.56 | 0.15 | -2.20 | -0.40 |
| Au | 0.82 | 0.57 | -0.07 | -2.84 | -0.50 |
| Pt | 1.08 | 0.50 | -0.82 | -1.03 | -0.84 |
| Ion Extraction after water decomposition (Fig. 1d) | | | | | |
| Al | -1.13 (-0.98) | 0.66 | 2.11 | 2.35 | 0.12 |
| Fe | 0.14 | 3.13 | 3.53 | 1.77 | 0.17 |
| Cu | 3.98 | 0.55 | 4.67 | 3.70 | 0.35 |
| Ag | 4.22 | 1.90 | 3.77 | 1.88 | 0.90 |
| Au | 4.75 | 2.66 | 3.93 | 2.34 | 0.53 |
| Pt | 5.02 | 2.48 | 5.97 | 4.38 | 1.27 |

## 5. Results of Calculations

Results of calculations of enthalpies of ion extractions and oxidation of metallic nanoparticles demonstrate the dependence of calculated energies from selected models. Qualitative results for all models except nanoparticles of 0.3 nm are, in general, the same sign and order of magnitude. Taking into account complete coverage of the surface by water (Fig. 3) does not provide significant changes in the energy costs of the studied processes. Decreasing of the energy costs of ion extraction in these cases is caused by additional coordination of extracted ions by water molecules. However, this decrease is much smaller than the energy costs of ion extraction because energies corresponding with metallic bonds are of orders higher than those energies of coordination bonds.

For the nanoparticles of the smallest size, results are in complete disagreement with the results obtained for other model systems. Thus, we can conclude that smallest nanoparticles (also called nanoclusters) have special chemical properties different from nanoparticles with sizes above 1 nm. Unfortunately, this mismatch between results for different models and the exclusion of the smallest nanoparticles from consideration, makes it impossible to reduce calculations of descriptors to a set of calculations for one system of the smallest size. On the other hand, we can obtain different descriptors from the analysis of the results for different models.

For all studied processes and materials, we can see significant dependence of the absolute values of energies from the chosen shape of the surface and from the type of object (surface, nanowire, nanoparticle). This result makes the development of descriptors more complicated, but permits us to evaluate the role of size and shape of nano-objects in biological activity. Additionally, we can see the difference in tendencies for two processes – ion extraction and water decomposition. This result permits us to evaluate two sources of possible toxicity (1) extraction of a single ion which further interacts with biological molecules, and (2) direct chemical interaction of nano-objects with biological molecules with their oxidation or reduction.

**Table2.** Descriptors (see in text) obtained by calculations based on the energies reported in Table1.

| Metal/Descriptor | Activity (A) | Size dependency (Sz) | Shape dependency (Sh) |
|---|---|---|---|
| Water decomposition (oxidation, Fig. 1c) | | | |
| Al | +2.48 | 1.04 | 0.78 |
| Fe | +2.77 | 0.03 | 0.58 |
| Cu | +1.66 | 2.43 | 0.65 |
| Ag | +2.20 | 2.78 | 0.03 |
| Au | +2.84 | 3.54 | 0.25 |
| Pt | +1.03 | 1.82 | 0.58 |
| Ion extraction (Fig. 1a, d) | | | |
| Al | +1.13 | 0.46 | 1.41 |
| Fe | +1.21 | 1.57 | 1.71 |
| Cu | -0.55 | 0.08 | 0.31 |
| Ag | -1.88 | 2.08 | 1.67 |
| Au | -2.25 | 2.46 | 0.51 |
| Pt | -2.48 | 0.43 | 1.60 |

## 6. Proposed descriptors

Obtained energy costs (values of enthalpy of reaction) from Table 1 for the selected model of nanoparticles cannot be directly used as descriptors of materials, but they can be united by simple calculations to a set of numerical values, which can be discussed as possible descriptors of the biological activity of studied systems.

## 6.1. Activity ($A_0$)

We propose the main descriptor is an activity ($A_0$) as minimal for all considered types of system values for a selected type of activity (ion extraction or oxidation – water dissociation) multiplied by negative one (-1). The cause of this multiplication is the comfort of perception of the descriptors. In the case of discussion about energetics, minimal energy costs of processes (or enthalpy of reaction) correspond with the most favorable process. Nevertheless, when we discuss activity, it is better to connect higher activity with the highest values of the descriptor. Note that ion extraction can also characterize the stability of the nanoparticles in aqueous media. Higher activity of ion extraction corresponds to the instability of nanoparticles in aqueous media.

For all nanoparticles considered in our work, values of activity for water decomposition are positive, corresponding to the high redox activity of metallic nanoparticles described in multiple works about the catalytic activity of these systems (see Refs. [12-14] and references therein). However, the main issue is the combination of redox activity with stability in aqueous media. This descriptor characterizes Pt-nanoparticles as safe for both – ion extraction and catalysis. Iron and aluminum are unstable because they have a small energy of ion extraction and oxidation. Copper and silver are metastable and can be a source of ions, but in the case of silver, the time of decomposition of nanoparticles in living tissues is much longer. All studied metals, other than Pt, demonstrate high activity in water decomposition which can be attributed to their catalytic activity in living tissues, but for unstable and metastable tissues, this activity is limited by the time of their survival inside bodies. In other words, the activity of Fe and Al nanoparticles is limited to the areas of incorporation in a body, copper and silver will penetrate deeper and act longer, and gold nanoparticles will be stable but chemically active in bodies for a longer time. Note: general activity should be discussed in combination with size dependency (see below) and for several materials, the decreasing of sizes can provide valuable increasing of activity, but this descriptor ($A_0$) can be used solely for discussing the activity of nanoparticles made from different materials.

**6.2. Size dependence (Sz)**

We define size dependence (Sz) as the magnitude of the difference between average energies of selected processes for (001) and (111) surfaces, and average energies for the same processes of nanoparticles of 0.6 nm multiplied by negative one (-1). A large value corresponds to significant size dependence and a small value with size-independence of activity of nanoparticles. From the results of these calculations (Table 2), we can see that catalytic activity of copper and platinum is size-dependent, and gold and silver nanoparticles are strongly size-dependent. These results are in qualitative agreement with experimental results reported in the literature.[12,13] In the case of ion extraction, only gold and silver demonstrate significant dependence of activity on size, which is also in qualitative agreement with experimental results.[39-42]

**6.3. Shape dependence (Sh)**

Nanoparticles of different shapes have surfaces similar to surfaces of different metals [43]. Based on this experimental evidence, we propose the magnitude of the difference between energies of the same process for (001) and (111) surfaces as a descriptor of shape dependence. Higher values correspond with higher shape dependence. The results of calculations demonstrate that the shape is more important for ion extraction than for water oxidation. Non-negligible value of size dependence for water oxidation on platinum surface is corresponding with experimentally detected dependence of catalytic activity of Pt nanoparticles from the shape.[44] The usage of this descriptor for the calculation of total activity is not clear at current stage of development of the concept. Perhaps after the calculation of this descriptor for other materials (oxides, carbon-based nanosystems), it will become clear.

## 6.4. Overall descriptor of activity

We can combine these three descriptors into a single equation. For size dependency, we should consider that activity increases with a reduction of radii, thus the contribution from size dependency is ~1/R. An increase in the size of nanoparticle provides the turn from various cluster-like structures from several atoms to bulk-like structures (see Fig. 2). In the other words, the coordination number of an atom on the surface of nanoparticles is different for the smallest size of nanoparticles, and almost the same as bulk-like structures for larger ones. Here, similar to size dependence, because nanoparticles became bulk-like with an increase of size, we multiply shape dependence coefficient by 1/R.

$$A^i = A^i_0 + Sz^i \ast a_{sz}/R + Sh^i \ast a_{sh}/R,$$

where $A^i$ – is the total activity for the selected type of activity $i$ (water decomposition or ion extraction), and R is the size of nano-objects in nm. The formula also contains the constants $a_{sz}$ and $a_{sh}$. Because size dependency of biological activity of nanoparticles was reported in multiple works, in contrast to shape dependency, we can conclude that $a_{sz} > a_{sh}$, and propose the value for these constants as 5 and 1, respectively. Thus the maximal contribution in activity from a nanoparticle 10 nm in size (the minimal size of realistic nanoparticles) is $Sz/2 + Sh/10$, or for the largest values of Sz and Sh from Table 2, the radii dependent contribution will be in order of 1.8 + 0.17, which is in the same order as $A_0$. For a nanoparticle of R = 100 nm, these contributions in total activity will be 0.18 + 0.017. Thus, we can conclude that for both types of activities (redox and ion extraction), the formula of overall descriptor:

$$A^i = A^i_0 + Sz^i * 5/R + Sh^i/R,$$

seems rather reasonable. Note that the discussed coefficient and the usage of $R^{-1}$ instead $R^{-a}$ is the first approach to the evaluation of biological activity of nanoparticles, and further, with the appearance of new experimental and theoretical data, could be corrected.

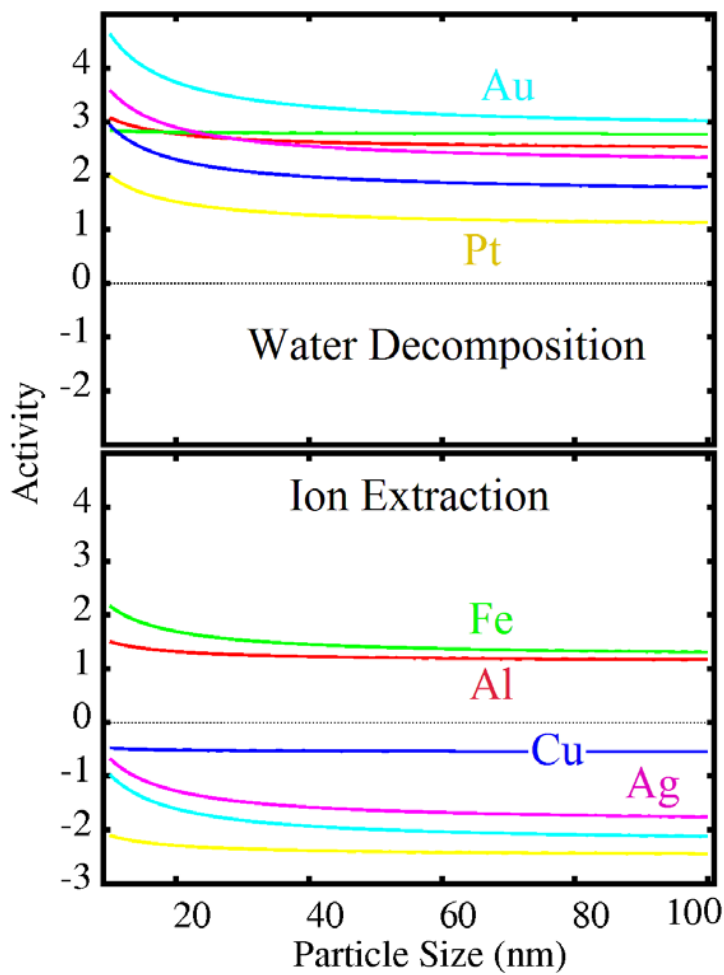

**Figure 4.** Activity as function of the particles size.

Visualization of descriptors as a function of nanoparticles size (Fig. 4) supports the discussion above about the properties of nanoparticles such as stability of Pt nanoparticles that supported by experimental data[45] in contrast to the instability of Al, Fe, and Cu nanoparticles

(these nanoparticles can be stabilized by surface coating, see for example Refs. [46,47]), the easy ion extraction of silver particles,[48] high stability, and catalytic activity of gold nanoparticles in agreement with experimental results (see for review Ref. [49]), and lower activity of Pt nanoparticles.[44]

**Table 3.** Calculated energy cost (in eV) of ion extraction and oxidation of aluminum (001) surface for various parameters of the computational model. The minimal set of parameters suitable for accurate enough calculations of the enthalpy of reactions is marked in bold.

| Number of Layers | 5 | | 3 | | | |
|---|---|---|---|---|---|---|
| Size of supercell in xy-plane (nm) | **1.46×1.14** | | | | 0.96×1.14 | |
| Mesh cutoff (Ry) | 360 | 50 | 360 | **50** | 360 | 50 |
| Ion extraction (Fig. 1b) | 4.87 | 4.83 | 4.74 | **4.82** | 4.92 | 5.19 |
| Water decomposition (oxidation, Fig. 1c) | 0.21 | 0.30 | 0.28 | **0.41** | -0.85 | -0.61 |
| Ion Extraction after water decomposition (Fig. 1d) | -1.13 | -0.94 | -1.32 | **-1.01** | 0.08 | 0.11 |

## 7. Minimization of computational costs

The main difficulty in using the discussed approach for all elements in the periodic table, as well as for oxides, alloys, and other systems, is the high computational costs for calculations of the processes over the surface of the slab (Fig. 1). There are several ways to minimize the computational costs without a noticeable change in the values of obtained descriptors. We varied some parameters of calculations or tried to decrease the number of atoms in the supercell. First, we decreased the k-point mesh from 3×3×1 to 1×1×1. The changes of the values of enthalpies of reactions were colossal. Thus, we could not save computational time this way. The next possible route was decreasing the k-point mesh. We performed step-by step variations of this parameter from 360 to 30 Ry and found that for values above 50 Ry, obtained enthalpies of reactions were almost the same (within 5%) as calculated for the higher value. Therefore, we can minimize computational cost by a decrease of cutoff energy. The next step is decreasing of the number of layers in the slab. We performed calculations for four and two layers, and the smallest supercell for minimal and maximal values of cutoff energy for the same set of k-point mesh. Results of calculations (Table 3) demonstrate that for our model, we can decrease not only the cutoff energy, but also number of the layers in the slab. Further decreasing the size of the supercell in the xy-plane provides results significantly different from results obtained for the full model. Therefore, for further calculations of descriptors we can use the minimal model which decreases computational costs by a factor of ten. This method for the test of the minimal model can be also applied for the development of similar computational descriptors of oxides and other nano-systems.

## 8. Conclusions

As a result of trial first-principles evaluation of chemical activity of metallic nanoparticles, we find that shape and size of the model system play an important role for both modelled processes (water decomposition and ion extraction in aqueous media), and calculations for various types of surfaces (different crystallographic surfaces, edges of nanowires and nanoparticles) are required for correct approximations. On the other hand, obtained results permit us to discuss chemical activity of nanoparticles, not only as a function of their chemical composition, but also as a function of their size. Results of evaluations of activities of various metallic nanoparticles (Al, Fe, Cu, Ag, Au, Pt) are in qualitative agreement with experimental evaluation of the stability and chemical activity of metallic nanoparticles. Additional calculations performed in this study demonstrate the possibility of significantly decreasing computational costs for obtained proposed numerical descriptors.

**Abbreviations**

QSAR - quantitative structure-activity relationship

SIESTA - Spanish initiative for electronic simulations with thousands of atoms

GGA-PBE - Generalized Gradient Approximation Perdew-Burke-Ernzerhof

**Founding Sources**

This work was supported by the Industrial Strategic Technology Development Program (10043929, Development of "User-friendly Nanosafety Prediction System"), funded by the Ministry of Trade, Industry & Energy (MOTIE) of Korea.

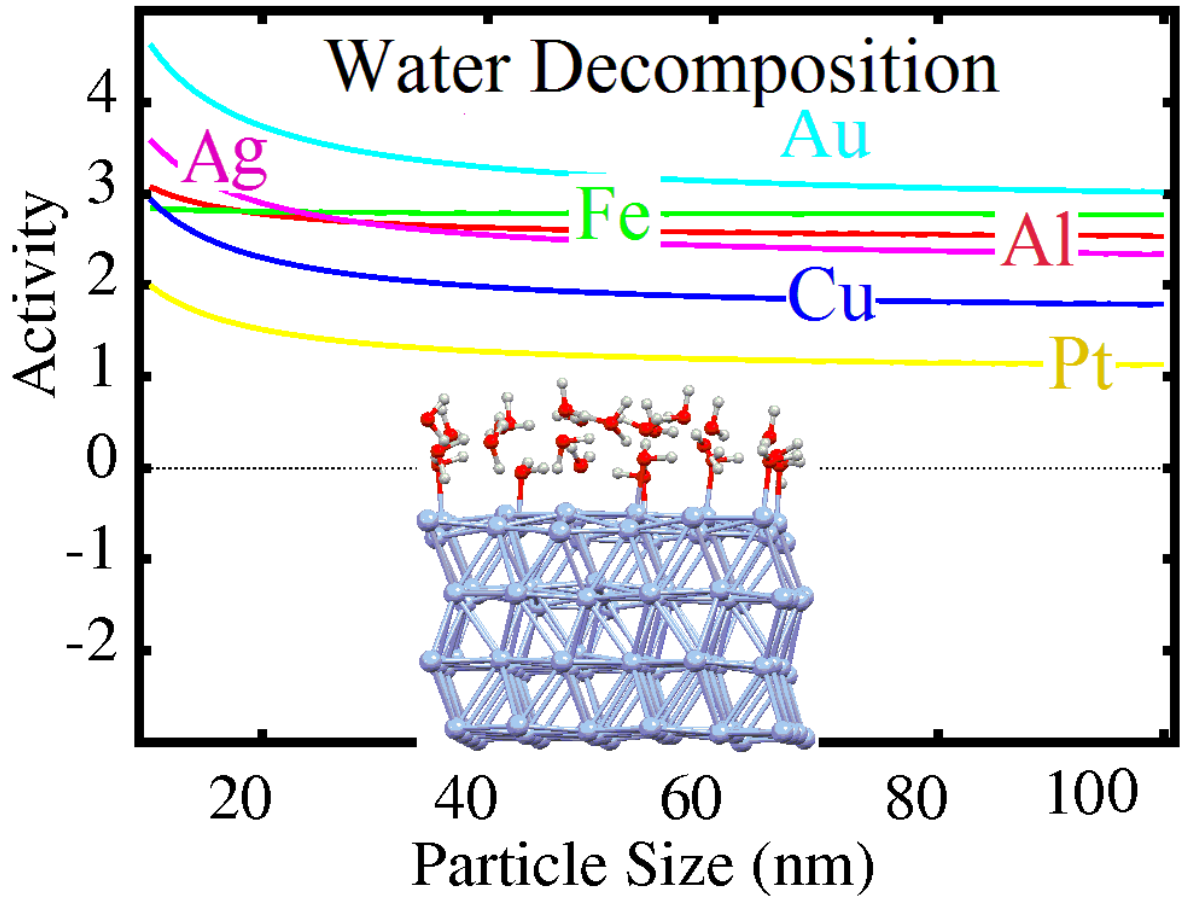

TOC graphics